\documentclass[aps,prb,reprint,superscriptaddress, nofootinbib,longbibliography]{revtex4-2}
\pdfoutput=1
\usepackage{graphicx}
\usepackage{amssymb}
\usepackage[linktoc=none]{hyperref}
\usepackage[dvipsnames]{xcolor}
\usepackage{dcolumn}
\usepackage[english]{babel}
\usepackage[utf8x]{inputenc}
\usepackage[T1]{fontenc}
\usepackage{eqnarray,amsmath,physics,bbold}
\usepackage{ dsfont }
\usepackage{appendix}
\usepackage{xcolor}
\usepackage{breqn}

\usepackage{hyperref}
    \hypersetup{
    colorlinks   = true,    
    urlcolor     = blue,    
    linkcolor    = blue,    
    citecolor    = red      
    }

\DeclareMathAlphabet{\mathpzc}{OT1}{pzc}{m}{it}

\newcommand{\AMSTERDAM}{Institute for Theoretical Physics Amsterdam,  University of Amsterdam, Science Park 904, 1098 XH Amsterdam, The Netherlands}

\newcommand{\IFWctqmat}{Institute for Theoretical Solid State Physics, IFW Dresden and W\"{u}rzburg-Dresden Cluster of Excellence ct.qmat, Helmholtzstrasse 20, 01069 Dresden, Germany}

\newcommand{\pd}{{\phantom{\dag}}}

\begin{document}
\title{Self-consistent surface superconductivity in time-reversal symmetric Weyl semimetals}

\author{Mattia Trama}
\email{m.trama@ifw-dresden.de}
\affiliation{\IFWctqmat}

\author{Viktor K\"onye}
\affiliation{\IFWctqmat}
\affiliation{\AMSTERDAM}

\author{Ion Cosma Fulga}
\affiliation{\IFWctqmat}

\author{Jeroen van den Brink}
\affiliation{\IFWctqmat}

\begin{abstract}
Weyl semimetals host topologically protected surface states, the so-called Fermi arcs, that have a penetration depth into the bulk that depends on surface-momentum, and diverges at the Weyl points. 
It has recently been observed in PtBi$_2$ that such Fermi arc states can become superconducting, with a critical temperature larger than that of the bulk.
Here we introduce a general variational method that captures the interplay between surface and bulk superconductivity, for any bulk Hamiltonian that harbors (topological) surface states with varying penetration depth. 
From the self-consistent solutions we establish that the surface state localization length of Weyl semimetals leads to characteristic features in the surface superconductivity, with a gap depending on surface momentum and a penetration length for the order parameter that is temperature-dependent due to competition with the bulk superconductivity.
\end{abstract}

\maketitle
The possibility of superconductivity at surfaces being enhanced compared to that of the bulk has a long history, and was considered already by Ginzburg and Kirzhnits 60~years ago~\cite{ginzburg1964superconductivity, ginzburg1964surface}. 
During the past couple of decades, this theoretical possibility found experimental confirmation with the observation of two-dimensional (2D) superconductivity, naturally realized in quasi-2D electronic gases at interfaces between oxides, the paradigm being the LaAlO$_3$/SrTiO$_3$ interface~\cite{ohtomo2004high, reyren2007superconducting,brun2016review,gariglio2009superconductivity}, or in ultrathin materials~\cite{zhang2010superconductivity,ye2012superconducting,cui2015multi,saito2016superconductivity,xi2016ising,costanzo2016gate}.

More recently, an interplay of bulk electronic topology and surface superconductivity has emerged.
Tunneling~\cite{schimmel2023high} and photoemission spectroscopy~\cite{kuibarov2024} have confirmed that the surface of trigonal PtBi$_2$ superconducts, with a critical temperature up to one order of magnitude larger than that of the bulk~\cite{veyrat2023berezinskii,shipunov2020polymorphic,zabala2024enhanced,vocaturo2024electronic}.
Interestingly, this material is a time-reversal (TR) invariant Weyl semimetal, and the surface states gapped out by superconductivity are its topological Fermi arcs~\cite{kuibarov2024,vocaturo2024electronic,hoffmann2024fermi}.

Given the potential implications of these findings for topological superconductivity, it is essential to understand what drives the pairing from which the surface superconductivity emerges as well as to address the large difference between surface and bulk critical temperatures. 
This requires a self-consistent theoretical framework that allows for an order parameter that depends on the position in the surface Brillouin zone (BZ). 

Here we introduce a variational approach that meets this challenge. 
A previous approach~\cite{nomani2023intrinsic} adopted an effective surface Green’s function method, with an order parameter included in the first surface layer. In contrast, we argue that the penetration of the surface states into the bulk is an essential component of the emergent superconductivity. Therefore, we introduce an independent method that generally allows 
for a self-consistent assessment of the superconducting gap opened both at the surface and in the bulk of the material. 

In the context of Weyl semimetals, given the recent observation of surface superconductivity, the application our method is of fundamental importance for two reasons.
First, their surface Fermi arcs are anomalous: they can only be realized at the surface of a 3D bulk, and not in any standalone 2D system~\cite{nomani2023intrinsic, hosur2012friedel}.
Second, their penetration depth is tied to their topology, given that their localization length diverges as they merge with the projections of the Weyl points on the surface BZ~\cite{haldane2014attachment, xu2015observation, matis2019surface, vocaturo2024electronic, chang2016signatures, huang2015weyl, sun2015topological, xu2015discovery}.

We apply our approach to a TR symmetric model for a type-I Weyl semimetal 
and find that the resulting surface superconductivity is not a 2D phenomenon, but rather the emerging manifestation of a full 3D phase transition on the surface.
We show that the numerical determination of the surface state wavefunctions is sufficient to deduce the self-consistent order parameter as a function of temperature. 
The enhanced density of states (DOS) leads to an increased surface superconductivity via an interplay between the spatial penetration of the surface states into the bulk and the spatially homogeneous order parameter opened in the bulk, which is determined by a conventional 3D Bardeen-Cooper-Schrieffer (BCS) theory \cite{bardeen1957theory}. 
The resulting gap opened at the surface has an intrinsic dependence on the quasi-momentum, which comes from the $k$-dependent penetration of the states enhancing or damping the surface order parameter. Such an effect can only be described by a self-consistent approach as the one presented here, making it essential for any study aiming at a description of the surface gap opening throughout the BZ.

{\textit{Semi-infinite superconductor}} --- 
We consider a TR symmetric Hamiltonian $H_0$ possessing some surface states when translational invariance is broken along the $\hat{z}$ direction. 
We consider a finite slab as a series of layers with an in-plane, 2D quasi-momentum $\mathbf{k}$, with Hamiltonian
\begin{equation}
H_0=\sum_{\mathbf{k},\alpha,\beta,\sigma,\sigma^\prime}h_{\sigma\sigma^\prime}^{\alpha\beta}(\mathbf{k})c^{\sigma\dagger}_{\mathbf{k}\alpha}c_{\mathbf{k}\beta}^{\sigma^\prime\vphantom{^\dag}}
\end{equation}
where $\sigma=\uparrow,\downarrow$ is the spin label and the indices $\alpha$ and $\beta$ contain the orbital ($l$) and the layer position ($z$) degrees of freedom of $h({\mathbf{k}})$. 
We will consider the Hamiltonian referred to a unit surface, so that $\sum_{\bf k}=\int d^2 {\bf k}/(2\pi)^2$.
Time-reversal symmetry allows for a local $s$-wave pairing so that the interaction Hamiltonian is $H=H_0+V$ with~\cite{bardeen1957theory, tinkham2004introduction}
\begin{equation}
    V=
    -U\sum_{\mathbf{r},\alpha} c^{\downarrow\dagger}_{\mathbf{r}\alpha} c^{\uparrow\dagger}_{\mathbf{r}\alpha}c^{\downarrow\vphantom{^\dag}}_{\mathbf{r}\alpha} c^{\uparrow\vphantom{^\dag}}_{\mathbf{r}\alpha},
\end{equation}
where we are now evaluating the creation and annihilation operator in real space as a function of the 2D coordinate $\mathbf{r}$ (the third coordinate, $z$, is included in $\alpha$), and $U$ is the interaction strength. 
For simplicity the pairing is assumed to be local on each layer and not mixing different orbitals. 
The order parameter is $\Delta_\alpha=\langle{c_{\mathbf{r}\alpha}^{\downarrow\vphantom{^\dag}}c_{\mathbf{r}\alpha}^{\uparrow\vphantom{^\dag}}}\rangle$ (independent of $\mathbf{r}$ due to translational invariance), and in mean field one finds, after going back to quasi-momentum space, an effective pairing term (per unit surface)
\begin{equation}
    V=-U\sum_{\mathbf{k},\alpha} \Delta^{*}_{\alpha} c^{\downarrow\vphantom{^\dag}}_{-\mathbf{k}\alpha}  c^{\uparrow\vphantom{^\dag}}_{\mathbf{k}\alpha} - U\sum_{\mathbf{k},\alpha}\Delta_{\alpha} c^{\downarrow\dagger}_{\mathbf{k}\alpha} c^{\uparrow\dagger}_{-\mathbf{k}\alpha} +U \sum_{\alpha} |\Delta_\alpha|^2,
\end{equation}
and the Hamiltonian is
\begin{dmath}
    H= H_{\text{BdG}}
    +\sum_{\mathbf{k}\in \frac{1}{2}\mathrm{BZ},\alpha} \left(h^{\alpha\alpha}_{\uparrow\uparrow}(-\mathbf{k})+h^{\alpha\alpha}_{\downarrow\downarrow}(-\mathbf{k})\right)+\sum_\alpha U|\Delta_{\alpha}|^2,
    \label{eq:BdGterms}
\end{dmath}
where the sum is over the half of the BZ and with its full expression provided in App.~\ref{app:details_bdg}.
One may neglect the trace of $h(-\mathbf{k})$ in Eq.~\eqref{eq:BdGterms} as it does not depend on $\Delta_\alpha$.
$H_{\text{BdG}}$ in the previous equation can be diagonalized in order to find the eigenvalues as
\begin{equation}
    H_{\text{BdG}}=\sum_{\mathbf{k}\in \frac{1}{2}\mathrm{BZ},i} \lambda_i(\mathbf{k}) \Gamma_{\mathbf{k}i}^\dagger\Gamma_{\mathbf{k}i}^\pd,
\end{equation}
where $\lambda_i$ are the eigenvalues of the system and $\Gamma_{\mathbf{k}i}$ are the annihilation operators of the states that, in principle, mix all the previous degrees of freedom and depend on $\Delta_\alpha$.
In order to find the self-consistent values of $\Delta_\alpha$ as a function of temperature $T$, one can minimize the free energy $F$ as a function of $\Delta_\alpha$, where
\begin{dmath}
    F=-k_B T \log(\mathcal{Z})=
    U\sum_{\alpha}|\Delta_\alpha|^2-\sum_{\mathbf{k}\in \frac{1}{2}\mathrm{BZ},i} k_{\rm B} T \log\left( 1+e^{-\beta\lambda_i(\mathbf{k})} \right),
    \label{eq:free_en}
\end{dmath}
where $\mathcal{Z}=\Tr{e^{-\beta H}}$ is the partition function of the system and $k_{\rm B}$ is the Boltzmann constant.
The minimization of Eq.~\eqref{eq:free_en} for all the values of $\Delta_\alpha$ is not trivial, since surface states will cause an enhanced order parameter close to the surface. 
To deal with the broken homogeneity, we introduce the Ansatz
\begin{equation}
    \Delta_\alpha=\Delta_0+\Delta_1 e^{- z/z_0},
    \label{eq:ansatz}
\end{equation}
where $\Delta_0$, $\Delta_1$ and $z_0$ are real and positive variational parameters. 
The exponential decay of the gap function reflects the exponential decay of the surface states; we comment later on the assumption of choosing all parameters real and positive.

For a semi-infinite system with $N\to \infty$ layers, electrons in the normal state can either form delocalized bulk states $\ket{b_{\mathbf{k}}^n}$ whose annihilation operator can be defined as $b^{n}_{\mathbf{k}}=\sum_{\sigma lz}\mathcal{U}^{n}_{\mathbf{k}\sigma lz} c_{\mathbf{k}\sigma lz}$ with $|\mathcal{U}^n_{\mathbf{k}\sigma lz}|\sim1/\sqrt{N}$ and eigenvalues $\xi^n_{\mathbf{k}}$, or they may form localized surface states $\ket{e_{\mathbf{k}}^n}$, whose associated annihilation operator is $e^n_{\mathbf{k}}=\sum_{\sigma lz}\tilde{\mathcal{U}}^n_{\mathbf{k}\sigma lz} c_{\mathbf{k}\sigma lz}$ with $|\tilde{\mathcal{U}}^n_{\mathbf{k}\sigma lz}|\sim \rho_{\mathbf{k}\sigma l}^{-z}$, $|\rho_{\mathbf{k}\sigma l}|<1$, and eigenvalues $\tilde{\xi}^n_{\mathbf{k}}$. 
We now rewrite $V$ in terms of the bulk and surface eigenstates, separating the terms depending on $\Delta_0$ and $\Delta_1 e^{-z/z_0}$. 
The homogeneous $\Delta_0$ does not mix bulk and surface states, since it is a constant, and basis changes maintain the property of not mixing different states. 
The inhomogeneous term contains $z$, hence the change of basis could cause mixing among eigenstates. 
However, for a semi-infinite system a decoupling is induced by the different spatial extensions of the states. 
When expressed in the new basis, the terms containing the projection over the bulk states vanishes since $\sum_z {\mathcal{U}}^n_{\mathbf{k}\sigma l z} {\mathcal{U}}^m_{-\mathbf{k}\sigma^\prime l^\prime z}e^{-z/z_0}\sim\sum_z e^{-z/z_0}/N \rightarrow 0$, similarly for the mixing of a bulk and a surface state, $\sum_z {\mathcal{U}}^n_{\mathbf{k}\sigma l z} \tilde{\mathcal{U}}^m_{-\mathbf{k}\sigma^\prime l^\prime z}e^{-z/z_0}\sim\sum_z e^{-z/z_0}/N \rightarrow 0$.
Only surface states are sizeably affected, since, $\sum_{\sigma\sigma^\prime l z}\Delta_1e^{-z/z_0}c_{\mathbf{k}\sigma lz}c_{-\mathbf{k}\sigma^\prime lz}=\sum_{\sigma\sigma^\prime lz}\sum_{nm} \Delta_1\left(\tilde{\mathcal{U}}^n_{\mathbf{k}\sigma lz}\tilde{\mathcal{U}}^m_{-\mathbf{k}\sigma^\prime l z}e^{-z/z_0}\right) e_{\mathbf{k}}^n e_{-\mathbf{k}}^m=\sum_{nm}\Delta_1 f_{\mathbf{k}}^n(z_0) e_{\mathbf{k}}^n e_{-\mathbf{k}}^n$ 
(the potential mixing matrix elements among different edge eigenstates can be neglected unless the Fermi surface includes a point where said eigenstates are nearly degenerate, as we discuss in detail in the App.~\ref{app:details_bdg}), where
\begin{equation}
    f^n_{\mathbf{k}}(z_0)=\sum_{\sigma lz} |\tilde{\mathcal{U}}^n_{\mathbf{k}\sigma lz}|^2 e^{-z/z_0}.
    \label{eq:penetration_function}
\end{equation}
In light of the above, one may separate the two blocks related to bulk and surface states with quasi-particles in the two blocks having energies $\lambda_n^b(\mathbf{k})=\pm\sqrt{(\xi^n_{\mathbf{k}})^2+U^2\Delta_0^2}$ for the bulk and $\lambda_n^e(\mathbf{k})=\pm\sqrt{(\tilde{\xi}^n_{\mathbf{k}})^2+U^2(\Delta_0+\Delta_1 f^n_{\mathbf{k}}(z_0))^2}$ for the surface. 
Replacing the Ansatz in Eq.~\eqref{eq:ansatz} and the eigenvalues in Eq.~\eqref{eq:free_en}, we find $F=N F_{\textrm{E}} + F_{\text{I}}$ where
\begin{dmath}
    F_{\textrm{E}} = U\Delta_0^2-\\ \frac{k_BT}{2}\int \frac{d^2 \mathbf{k}\; dk_z}{8\pi^3} \sum_{i} \log\left(1+e^{-\beta \Lambda_i^b(\mathbf{k},k_z)}\right),
    \label{eq:free_en_explicit_extensive}
\end{dmath}
\begin{dmath}
    F_{\textrm{I}} = 2U\frac{\Delta_0\Delta_1}{1-e^{-1/z_0}}+U\frac{\Delta_1^2}{1-e^{-2/z_0}}-\frac{k_B T}{2}\int \frac{d^2 \mathbf{k}}{4\pi^2}\sum_n \log\left(1+e^{-\beta\lambda_n^e(\mathbf{k})}\right),
    \label{eq:free_en_explicit_intensive}
\end{dmath}
where in $F_{\textrm{E}}$ the sum over the layers for the bulk states is converted as usual into an integral over the $k_z$ quasi-momentum, having introduced the 3D bulk eigenvalues $\Lambda_i^b(\mathbf{k},k_z)$.
The free energy separates into an extensive portion $N F_{\textrm{E}}$, growing with the number of layers $N$, and an intensive portion $F_{\textrm{I}}$ dominated by surface physics. 
The extensive terms reproduce the conventional BCS free energy depending only on $\Delta_0$ and the bulk states. 
Minimizing the extensive free energy reproduces $\Delta_0$ from bulk BCS theory. 
The intensive free energy is now affected by the bulk and surface states in two ways: 
\textit{i}) due to the mixing term $\Delta_0 \Delta_1$ and the impact of $\Delta_0$ on the eigenvalues of surface states, bulk superconductivity in general affects the surface superconductivity; 
\textit{ii}) the extension of the surface states into the first layers of the bulk, naturally encapsulated in the function $f^n_{\mathbf{k}}(z_0)$, renormalizes the effective order parameter that each surface state is subjected to. 
The function $f^n_{\mathbf{k}}(z_0)$ is always less than $1$, and is higher for well-confined surface states, which therefore for a fixed interaction strength are affected more strongly by the superconducting instability.
\\Notice that the use of short-range interaction implies a summation over momentum states even far from the Fermi surface. However, this does not signal a logical inconsistency in the theory, since the dependence on deep Fermi sea states is only logarithmic, and—as is standard in weak-coupling treatments—it is removed by renormalizing the coupling constant to match physical observables (see, e.g., Ref.~\cite{gor1961contribution}). In this sense, our approach is fully consistent with the weak-coupling limit.

The method described above has the advantage of disentangling the bulk from the surface problem, emphasising at the same time that the bulk order parameter heavily influences the value of the surface order parameter. 
Moreover, it suggests an operative way to determine the surface properties of real systems.
In fact, one can infer the strength of interaction $U$ from the critical temperature in the bulk. 
Assuming that the interaction strength is the same also at the surface, the surface physics is thus entirely determined. 
Should specific interaction channels be modified  at the surface, one may include them by introducing an inhomogeneous interaction strength. 

\begin{figure}
\centering
\includegraphics[width=0.49\textwidth, trim={0.6cm 0 0.cm 0}]{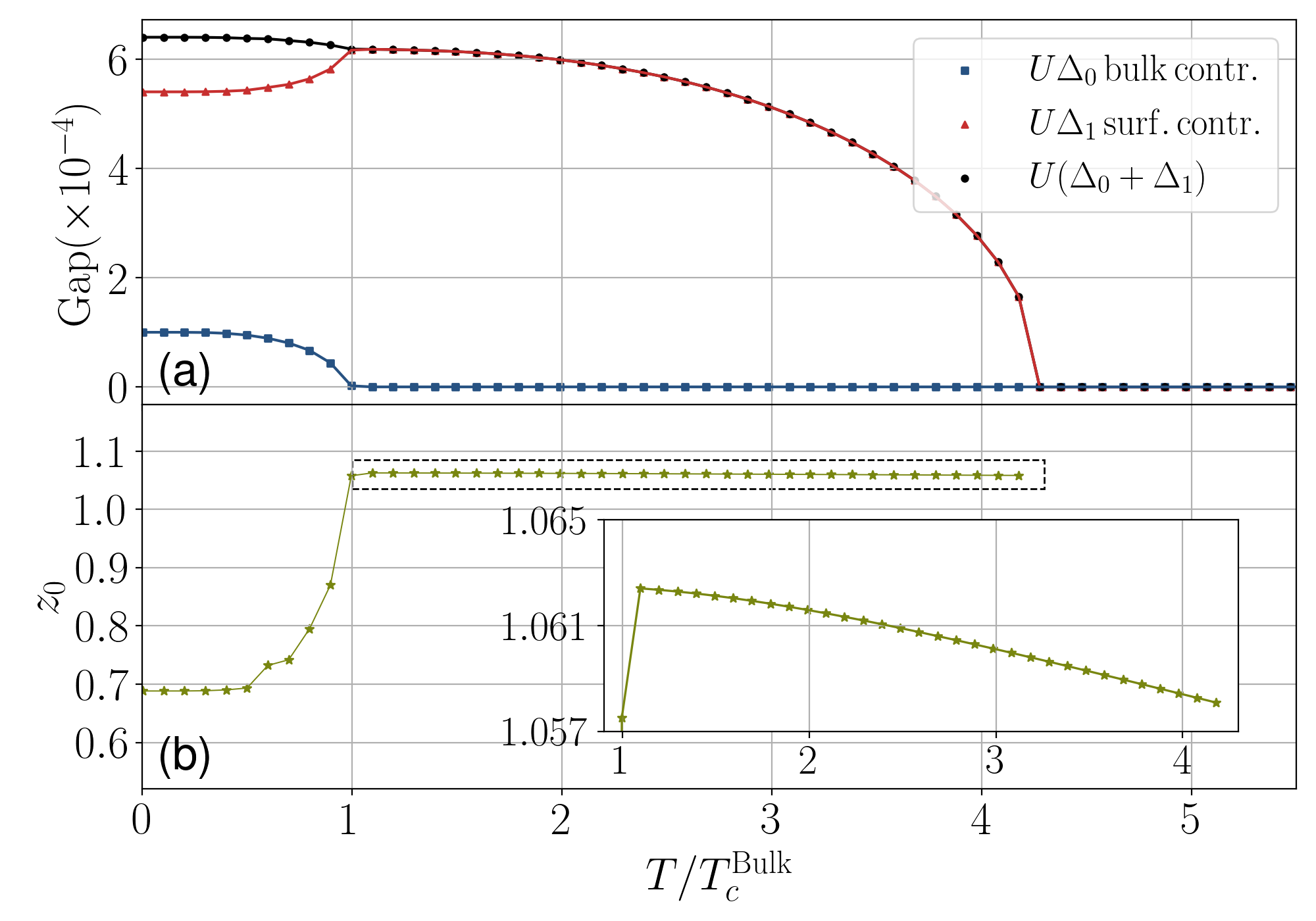}
\\
\includegraphics[width=0.49\textwidth, trim={0.6cm 1.3cm 0.0cm 0.3cm}]{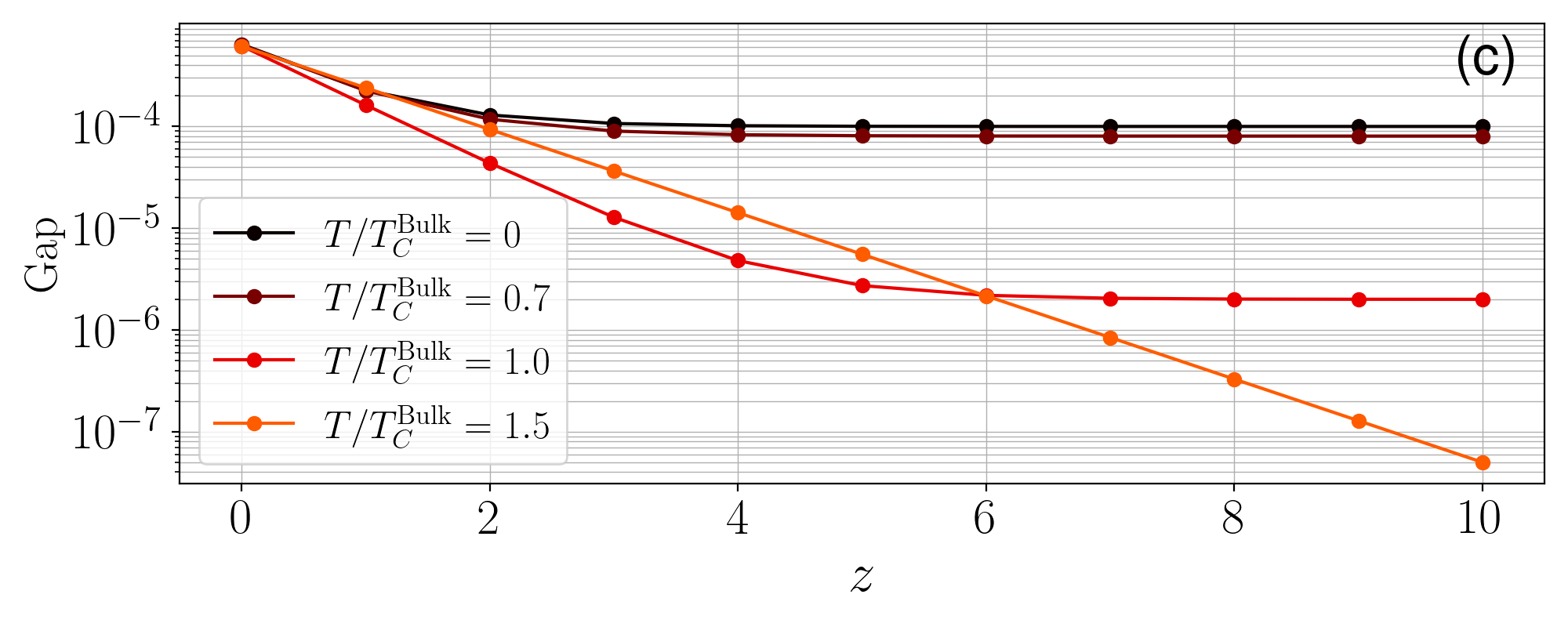}
\caption{
(a) Temperature dependence of the bulk and surface order parameters. Blue squares, red triangles, and black points correspond to the superconducting gap for bulk ($g_0=U\Delta_0$), surface ($g_1=U\Delta_1$) and  total order parameter of the first layer [$g_s=U(\Delta_0+\Delta_1)$], respectively. 
(b) Penetration depth $z_0$ of the surface order parameter as a function of the temperature. 
The inset details of the behavior of $1/z_0$ when the bulk order parameter has already vanished, showing that in this regime there is an opposite tendency of increasing penetration depth with increasing temperature. 
Temperature is scaled to the bulk critical temperature $T_c^{\textrm{Bulk}}$. (c) Total gap $U(\Delta_0+\Delta_1 e^{z/z_0})$ as a function of layer position, for different temperatures. 
The curve for $T/T_C^{\textrm{Bulk}}=1.0$ approaches a finite value since $T\rightarrow T_C^{\textrm{Bulk}}$ from left.
}
\label{fig:self_cons_mu_-0.03}
\end{figure}

\textit{Surface superconductivity in a Weyl semimetal} ---
We demonstrate our approach on a generic TR symmetric model for a Weyl semimetal proposed in Refs.~\cite{lau2017generic,kourtis2016universal} (in the App.~\ref{app:analytical}, we also confirm our numerical approach by applying it to a simpler, analytically-solvable model)
\begin{dmath}
    h(\mathbf{k},k_z)=a\left(\sin k_y \tau_y \sigma_x + \sin k_z \tau_z \sigma_0\right)+
    \beta\tau_z\sigma_z + d\tau_z \sigma_x +\left[t\cos k_x+2b\left(2-\cos k_y -\cos k_z\right)\right]\tau_x\sigma_0 +\alpha \sin k_z \tau_y \sigma_z +\lambda \sin k_x \tau_0 \sigma_y-\mu\tau_0\sigma_0,
    \label{eq:model_1}
\end{dmath}
where $\sigma_i$ and $\tau_i$ are the Pauli matrices along the direction $i$ respectively for the spin and orbital degrees of freedom, and the subscript $0$ refers to the identity in their subspace. Terms in $\lambda$, $d$, and $\beta$ break the inversion symmetry, while TR symmetry is always preserved.
Fermi arcs appear when translational invariance is broken along the $z$ axis. 
Bulk and surface spectral functions and the Fermi surface of this model for a benchmark choice of parameters is shown in the App.~\ref{app:band_structure}.  

We build a finite slab of $40$~layers, and numerically diagonalize it in the normal state. 
We separate the surface states based on their penetration in the bulk (specifically, such that 
$\sum_{\sigma l,z=0}^{z=5} \abs{ {\tilde{\mathcal{U}}^n_{\mathbf{k}\sigma lz}}}^2\geq0.7$).
For a $4\times4$ Hamiltonian, we expect only $2$ surface states 
on the top layer and $2$ on the bottom layer. 
For each of them, we determine the function $f^n_{\mathbf{k}}(z_0)$ using Eq.~\eqref{eq:penetration_function}, for $z_0\in[0.1,10]$.

\begin{figure}
\centering
\includegraphics[width=0.49\textwidth,trim={1.6cm 0.4cm 2.cm 0.4cm}]{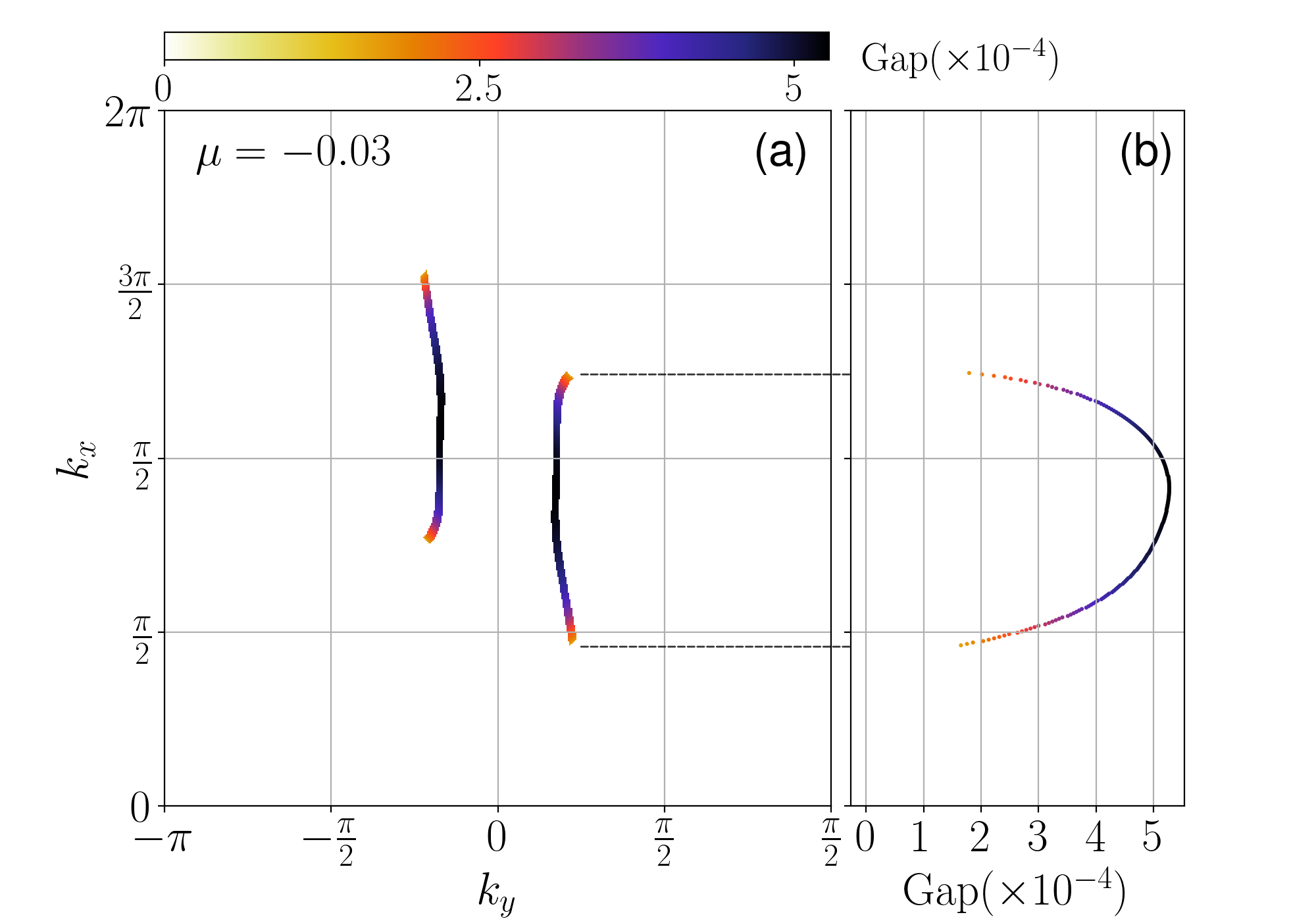}
\caption{
(a) Fermi arcs with their gap $U\left[\Delta_0+\Delta_1 f(z_0)\right]$ (color scale) at $\mu=-0.03$ in the 2D BZ spanned by $k_x$ and $k_y$. 
(b) Momentum dependence of the energy gap along the Fermi arc. The highlighted Fermi surface has been found by selecting an energy window of $2\delta w=5\cdot10^{-2}$ centered on the Fermi arcs, i.e. by computing $U\left[\Delta_0+\Delta_1 f(z_0)\right]\Theta(\mu-\tilde{\xi}+\delta w)\Theta(-\mu+\tilde{\xi}+\delta w)$, with $\Theta(\cdot)$ the Heaviside function.}
\label{fig:gap_k}
\end{figure}

We first minimize the bulk free energy to obtain the conventional BCS results for the bulk order parameter as a function of temperature. 
We use this as a scale, fixing the interaction strength $U$ to the value required to open a gap in the bulk $g_0=U\Delta_0=10^{-4}$ at zero temperature. 
With this value of $U$, and the self-consistent order parameter $\Delta_0(T)$, we identify $\Delta_1(T)$ and $z_0(T)$ by minimizing the surface free energy. 
Figure~\ref{fig:self_cons_mu_-0.03} shows the results for chemical potential $\mu=-0.03$ (the role of chemical potential is further discussed in App.~\ref{app:fit_dos}). 
We find a non-vanishing surface order parameter $\Delta_1$, which is in fact higher than $\Delta_0$ at zero temperature, due to the higher local DOS at the surface. 
The penetration depth $z_0$ has a non-trivial temperature dependence, increasing with the decreasing bulk order parameter $\Delta_0$. 
We interpret this as a competition between bulk and surface superconductivity: with decreasing $\Delta_0$, electrons at the surface tend to compensate for the reduction by extending the surface order parameter into the bulk, an effect that can only be captured from the 3D approach. Indeed, its origin can be traced to the mixed product term $\Delta_0 \Delta_1$ in Eq.~\eqref{eq:free_en_explicit_intensive}: a non-vanishing bulk parameter $\Delta_0$ introduces an additional energy cost for $\Delta_1$, disfavoring surface superconductivity.
The explicit spatial dependence of the order parameter is shown in Fig.~\ref{fig:self_cons_mu_-0.03} for varying temperatures. As the temperature is increased above the critical value for the bulk, the order parameter far from the interface decreases as expected, yet in the very first layers there is an overall spreading, due to the increasing value of $z_0$.
Finally, the surface states contribute to the free energy weighted by the factor $f_{\mathbf{k}}(z_0)$, measuring the degree of penetration of each state into the material. This weighting factor depends on $\bf{k}$, so that a momentum dependence of the surface-related gap emerges,  
despite the absence of long-range components of the electron-electron interaction. 
This is visible in Fig.~\ref{fig:gap_k}, from which one sees that the gap at the first layer is very strongly $\mathbf{k}$-dependent. 

\textit{Summary and conclusions} --- We considered superconductivity at surfaces of TR invariant materials which possess surface states, where one may expect the modified density of states in the vicinity of the surface to cause superconductivity with a critical temperature different from the bulk one. 
By introducing a variational method we tackled this problem, that takes into account the interplay between surface and bulk states in the formation of a superconducting order parameter.
It enables us to quantitatively determine the extent to which surface superconductivity penetrates through multiple layers away from the surface, and is able to capture its emergence out of the underlying 3D model.
We have applied this approach on a generic Weyl Hamiltonian, as a prototype of a system for which an effective, analytical surface Hamiltonian does not exist, and the emerging surface states are found numerically. The method can be directly extended to more complex Hamiltonians with a larger number of degrees of freedom. 
Despite the lack of a long-range component, we find that the pairing interactions still induces the opening of a gap on the surface which varies across the BZ, where the variability is induced by the changing penetration depth of the surface states into the bulk, a feature that clearly can only be captured by a 3D approach.  We also showed that bulk superconductivity at sufficiently low temperatures competes with the emerging surface superconductivity. The results suggest a route towards the explanation of the recent observed surface superconductivity in PtBi$_2$ without a corresponding bulk superconductivity, and provides the tools to investigate this emerging behavior with quantitative accuracy.

Interestingly the presence of an inhomogeneous order parameter, which forms the basis of our modeling, opens intriguing avenues for further exploration (for example in App.~\ref{app:more_ansatzes} we have tested more complex patterns for the spatial dependence of the order parameter). Here we have explored the starting point that the inhomogeneous part of the order parameter is real and with the same sign as the bulk order parameter.  
However, the inhomogeneous component of the order parameter may in principle have its own phase, which may also be space-dependent. Our variational approach is particularly suited for such generalizations, since it only requires minimization with respect to the space-varying phase as well as the amplitude of the order parameter. 

We adopt a short-range interaction, as in the original BCS treatment. A potential longer range component may definitely be present, in which case however the method we have introduced can be applied in the same way, with the only additional complexity of a dependence of the order parameter on the wavevector. This should cause no conceptual complication to our method, and we plan to consider this case in future work.
Overall, the method we propose allows for a systematic and flexible investigation of the penetration of the order parameter inside the material. Our framework relies on a controlled approximation -- its convergence can be tested by simply extending the variational ansatz -- and can be easily generalized to arbitrary models, offering the possibility of a systematic characterization of surface superconductivity. Such a task, which until recently may have been of mainly theoretical interest, has become extremely relevant given the recent observations of surface superconducting states.

\section*{Acknowledgments}
We thank Damiano Fiorillo, Riccardo Vocaturo, Roberta Citro and Carmine Antonio Perroni for enlightening discussions, and Ulrike Nitzsche for technical assistance.
M.T. acknowledges financial support from "Fondazione Angelo Della Riccia".
We acknowledge support from the Deutsche Forschungsgemeinschaft (DFG, German Research Foundation) under Germany’s Excellence Strategy through the W\"{u}rzburg-Dresden Cluster of Excellence on Complexity and
Topology in Quantum Matter – ct.qmat (EXC 2147, project-ids 390858490 and 392019).

\appendix

\section{Details of the Bogoliubov-de-Gennes Hamiltonian}\label{app:details_bdg}
In this section we give the details of the BdG Hamiltonian presented in the main text and the explicit form of the diagonalized Hamiltonian.
By referring to the notation introduced in the main text, and by using the particle-hole basis as 
\begin{equation}
    \gamma_{\mathbf{k}\alpha}=\begin{pmatrix}
        c_{-\mathbf{k}\alpha}^{\uparrow\dagger}\\
        c_{-\mathbf{k}\alpha}^{\downarrow\dagger}\\
        c_{\mathbf{k}\alpha}^{\downarrow\vphantom{^\dag}}\\
        c_{\mathbf{k}\alpha}^{\uparrow\vphantom{^\dag}}
    \end{pmatrix},
\end{equation}
the full Hamiltonian $H=H_0+V$ can be written in a mean field approach in the following way
\begin{dmath}
    H= H_{\text{BdG}}
   +\sum_\alpha U|\Delta_{\alpha}|^2,
   \label{eq:BdGterms_app}
\end{dmath}
where 
\begin{widetext}
\begin{dmath}
H_{\text{BdG}}=\sum_{\mathbf{k}\in \frac{1}{2}\mathrm{BZ}, \alpha, \beta}\gamma_{-\mathbf{k}\alpha}^\dagger
    \begin{pmatrix}
    h_{\uparrow\uparrow}^{\alpha\beta} (\mathbf{k}) & h_{\uparrow\downarrow}^{\alpha\beta}(\mathbf{k}) & 0 & -U\Delta_\alpha\delta_{\alpha\beta}\\
    h_{\downarrow\uparrow}^{\alpha\beta}(\mathbf{k}) & h_{\downarrow\downarrow}^{\alpha\beta} (\mathbf{k}) & U\Delta_{\alpha}\delta_{\alpha\beta} & 0\\
    0 & U\Delta^*_\alpha\delta_{\alpha\beta} & -h_{\downarrow\downarrow}^{\beta\alpha} (-\mathbf{k}) & -h_{\downarrow\uparrow}^{\beta\alpha} (-\mathbf{k})\\
    -U\Delta^*_\alpha\delta_{\alpha\beta} & 0 & -h_{\uparrow\downarrow}^{\beta\alpha} (-\mathbf{k}) & -h_{\uparrow\uparrow}^{\beta\alpha} (-\mathbf{k})
    \end{pmatrix}
    \gamma_{\mathbf{k}\beta}\vphantom{^\dag}.
\label{eq:BdG_Ham_app}
\end{dmath}
\end{widetext}
and having neglected the trace of $h(-\mathbf{k})$ in Eq.~\eqref{eq:BdGterms_app} which does not depend on $\Delta_\alpha$.
$H_{\text{BdG}}$ in the previous equation can be diagonalized in order to find the eigenvalues as
\begin{equation}
    H_{\text{BdG}}=\sum_{\mathbf{k}>0,i} \lambda_i(\mathbf{k}) \Gamma_{\mathbf{k}i}^\dagger\Gamma_{\mathbf{k}i}^\pd,
    \label{eq:diag_bdg}
\end{equation}
where $\lambda_i$ are the eigenvalues of the system and $\Gamma_{\mathbf{k}i}$ are the annihilation operators of the states that, in principle, mix all the previous degrees of freedom and depend on $\Delta_\alpha$.
However, as we showed in the main text, the diagonalization can be performed easily by using the variational ansatz $\Delta_\alpha=\Delta_0+\Delta_1 e^{-z/z_0}$. Due to the separation among the surface and bulk states, the Hamiltonian~\eqref{eq:diag_bdg} admits the following block form in the basis of the edge and bulk eigenstates of the normal Hamiltonian, respectively $\ket{e_{\mathbf{k}}^n}$ and $\ket{b_{\mathbf{k}}^n}$, corresponding to the annihilation operators $e_{\mathbf{k}}^{n}$ and $b_{\mathbf{k}}^{n}$ in second quantization. Using the basis 
\begin{equation}
    \tilde{\gamma}_{\mathbf{k}n}=\begin{pmatrix}
        e_{\mathbf{k}}^{n\dagger}\\
        e_{-\mathbf{k}}^{n\vphantom{^\dag}}\\
        b_{\mathbf{k}}^{n\dagger}\\
        b_{-\mathbf{k}}^{n\vphantom{^\dag}}
    \end{pmatrix},
\end{equation}
the BdG Hamiltonian reads

\begin{widetext}
    \begin{equation}
        H_{\text{BdG}}=\sum_{\mathbf{k}\in \frac{1}{2}\mathrm{BZ}, n}\tilde{\gamma}_{-\mathbf{k}n}^\dagger
            \begin{pmatrix}
            \tilde{\xi}_{\mathbf{k}}^n & -U\left(\Delta_0+\Delta_{1}f_{\mathbf{k}}^n(z_0)\right) & 0 & 0\\
           U\left(\Delta_0+\Delta_{1}f_{\mathbf{k}}^n(z_0)\right) & -\tilde{\xi}_{-\mathbf{k}}^n & 0 & 0\\
            0 & 0 & \xi_{\mathbf{k}}^n & -U\Delta_0\\
            0 & 0 & U\Delta_0 & -\xi_{-\mathbf{k}}^n
            \end{pmatrix}
            \tilde{\gamma}_{\mathbf{k}n}\vphantom{^\dag}.
    \end{equation}
\end{widetext}
where the matrix elements of the off-diagonal block terms belonging to the subspace of the edge states have been computed as 
\begin{widetext}
\begin{equation}
\sum_{\sigma\sigma^\prime l z}\Delta_1e^{-z/z_0}c_{\mathbf{k}\sigma lz}c_{-\mathbf{k}\sigma^\prime lz}=
\sum_{\sigma\sigma^\prime lz}\sum_{nm} \Delta_1\left(\tilde{\mathcal{U}}^n_{\mathbf{k}\sigma lz}\tilde{\mathcal{U}}^m_{-\mathbf{k}\sigma^\prime l z}e^{-z/z_0}\right) e_{\mathbf{k}}^n e_{-\mathbf{k}}^m=
\sum_{n}\Delta_1 f_{\mathbf{k}}^n(z_0) e_{\mathbf{k}}^n e_{-\mathbf{k}}^n,
\end{equation} 
\end{widetext}

with
\begin{equation}
    f^n_{\mathbf{k}}(z_0)=\sum_{\sigma lz} |\tilde{\mathcal{U}}^n_{\mathbf{k}\sigma lz}|^2 e^{-z/z_0}.
    \label{eq:penetration_function_supp}
\end{equation}
having used the relation $\tilde{\mathcal{U}}_{-\mathbf{k}\sigma^\prime l}^{n}=\left(\tilde{\mathcal{U}}_{\vphantom{-}\mathbf{k}\sigma^{\vphantom{\prime} } l}^{n}\right)^*$.
The potential mixing among different edge states, induced by matrix elements of the form 
$\sum_{nm}\Delta_1 f_{\mathbf{k}}^{nm}(z_0) e_{\mathbf{k}}^n e_{-\mathbf{k}}^m$, with $f^{nm}_{\mathbf{k}}(z_0)=\sum_{\sigma lz} \tilde{\mathcal{U}}^n_{\mathbf{k}\sigma lz} (\tilde{\mathcal{U}}^m_{\mathbf{k}\sigma lz})^* e^{-z/z_0}$, 
would be relevant only if the chemical potential intersects both eigenstates simultaneously at the same value of $\bf k$, i.e. if the Fermi surface includes a point of degeneracy among the two eigenstates. This is because in the limit $U \ll \mu$, only states within a small energy interval of order $U$ from the Fermi surface are affected by the superconducting coupling, so only if the two states are close in energy their mixing could induce any consequence. Mathematically, unless $|\tilde{\xi}^n_{\bf k}|-|\tilde{\xi}^m_{\bf k}|\sim U$, a potential off-diagonal term of order $U$ would negligibly affect the eigenstates of the BdG Hamiltonian. For the regime we study here, this is never the case; it is straightforward to extend our method to include the off-diagonal mixing of edge states if the Fermi surface includes a degeneracy point.
The eigenvalues of the blocks therefore are $\lambda_n^b(\mathbf{k})=\pm\sqrt{(\xi^n_{\mathbf{k}})^2+U^2\Delta_0^2}$ for the bulk and $\lambda_n^e(\mathbf{k})=\pm\sqrt{(\tilde{\xi}^n_{\mathbf{k}})^2+U^2(\Delta_0+\Delta_1 f^n_{\mathbf{k}}(z_0))^2}$ for the surface. We note that we rewrite the bulk eigenvalues by explicitly taking into account the out-of-plane momentum $k_z$ dependency included in the label $n$, so that we can define $\Lambda_n^b(\mathbf{k},k_z)=\pm\sqrt{(\Xi^n_{\mathbf{k},k_z})^2+U^2\Delta_0^2}$, where $\Xi^n_{\mathbf{k},k_z}$ are the single particle bulk eigenvalues for which we made explicit $k_z$.

\section{Electronic band structure in the normal state}\label{app:band_structure}

\begin{figure}
    \centering
    \includegraphics[width=0.49\textwidth]{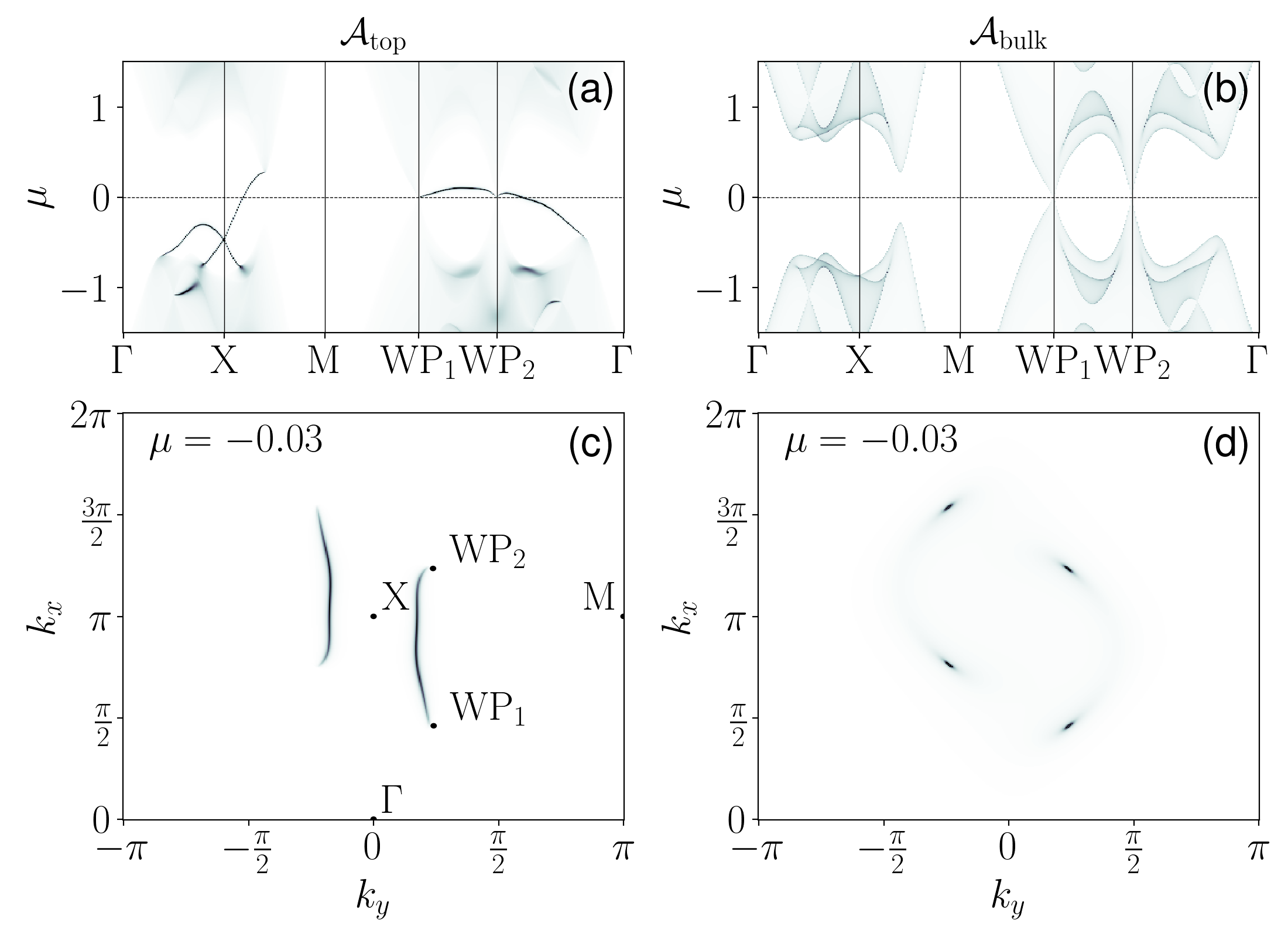}
    \caption{
    (a) Spectral function $\mathcal{A}_i$ of the top surface and (b) of the bulk of Hamiltonian in Eq.~(11) of the main text. The color maps of (a) and (b) are the same, and have been cut off at $\mathcal{A}^{\textrm{max}}=10$ to improve the contrast.  (c) Spectral function of the surface and (d) of the bulk for a benchmark choice of the chemical potential $\mu=-0.03$.  The color maps of the two plots have been chosen to enhance the contrast. The parameters of the Hamiltonian are $a=b=1$, $t=1.5$, $\alpha=0.9$, $d=0.2$, $\lambda=0.8$, $\beta=0.7$.}
    \label{fig:spect_function}
\end{figure}

The model presented in the main text exhibits Weyl points with a characteristic structure in the Brillouin zone (BZ). In this section, we show explicitly the band structure of the Hamiltonian in the normal state, i.e. for vanishing interaction strength $U=0$.

Fig.~\ref{fig:spect_function} shows the band structure and the spectral function for the Hamiltonian. To obtain the latter, we define a bare Green's function $\mathcal{G}^{-1}=h(\mathbf{k})+i\delta$, and use it to find the exact surface ($\mathcal{G}_{\text{top}}=\mathcal{G}_0$) and the bulk ($\mathcal{G}_{\text{bulk}}=\mathcal{G}_{N/2}$) Green's functions using the iterative method of Ref.~\cite{sancho1985highly}. The imaginary part of this Green function is connected to the spectral density as $\mathcal{A}_i=-\frac{1}{\pi}\text{Im} \Tr{\mathcal{G}_i}$.

\section{Density of states and order parameter dependence}\label{app:fit_dos}

\begin{figure}
\centering
\includegraphics[width=0.49\textwidth, trim={0 0 0 1.5cm}]{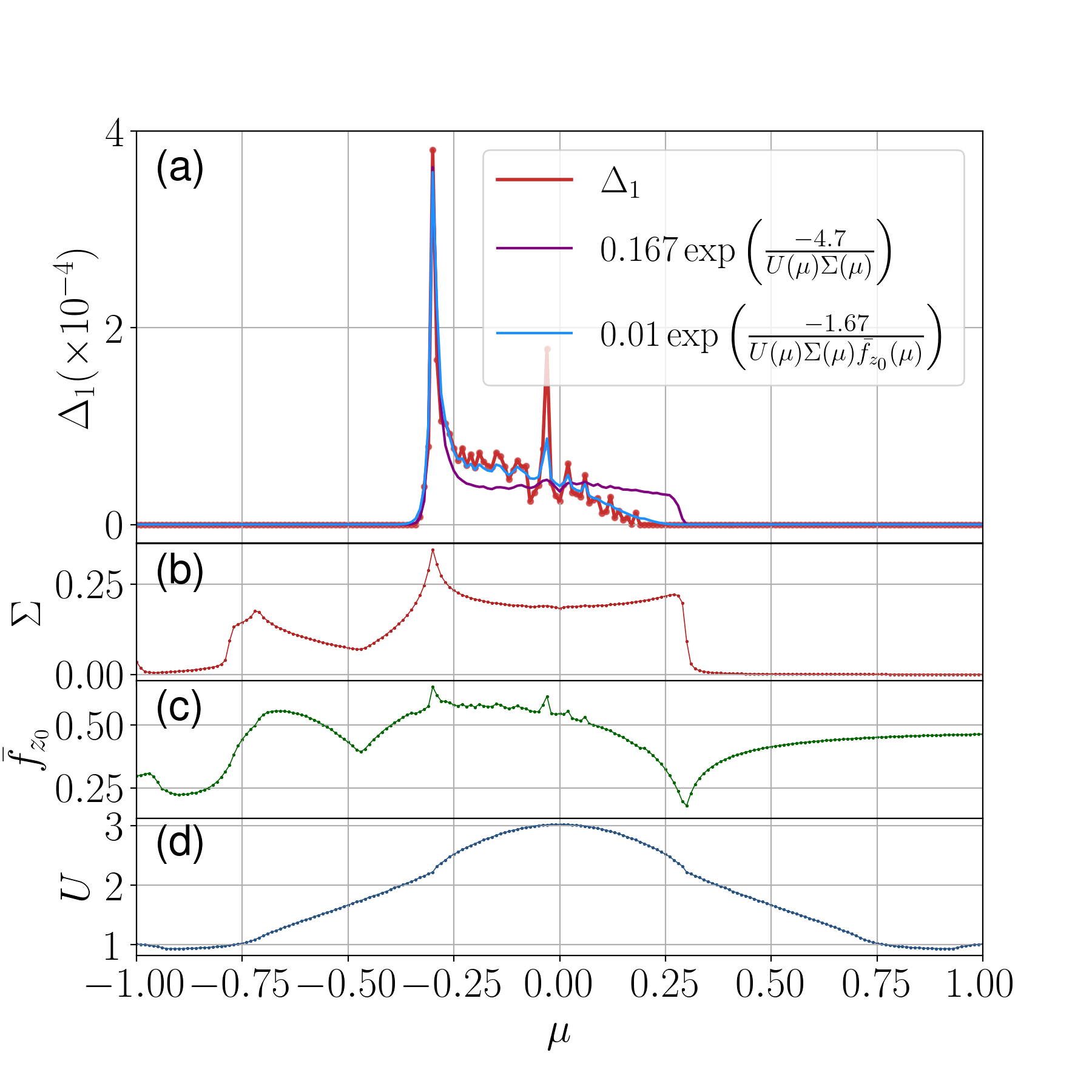}
\caption{
Chemical potential analysis. 
(a) $\Delta_1$ as a function of the chemical potential and fit of the BCS expression of the order parameters. 
(b) Surface density of states. 
(c) Average of the penetration function $f_{z_0}$ according to the self-consistent value of $z_0$ which depends on $\mu$. 
(d) Value of $U$ which opens a gap in the bulk of $U\Delta_0=10^{-4}$.
}
\label{fig:density_of_states_function}
\end{figure}

In this section we discuss the role of the chemical potential ($\mu$) and the surface density of states ($\Sigma$) in determining the surface contribution to the order parameter. 
According to the BCS theory~\cite{ANGHEL201674}, the bulk order parameter follows the equation
\begin{equation}
    \Delta_0=2\hbar \omega e^{-\frac{1}{U \Sigma_0}},
    \label{eq:bcs_ord_par}
\end{equation}
where $2\hbar\omega$ is the bandwidth of the single particle spectrum centered on $\mu$, for a system whose density of states $\Sigma_0$ is almost constant in that interval. 
We want to test how such an expression qualitatively fares with the description of the surface contribution. 
In Fig.~\ref{fig:density_of_states_function}(a) we show the self-consistent values of $\Delta_1$ as a function of the chemical potential, constraining the value of the bulk gap $g_0=U\Delta_0=10^{-4}$. In Fig.~\ref{fig:density_of_states_function}(b) we show the value of the surface local DOS computed as
\begin{equation}
    \Sigma(\mu)=\int_{BZ} \frac{d^2 \mathbf{k}}{4\pi^2} \mathcal{L}_{\delta\tau}(\tilde{\xi}_{\mathbf{k}}^n-\mu),
\end{equation}
where we highlighted the chemical potential dependence on the surface eigenvalues and we defined $\mathcal{L}(x)=\frac{1}{\pi}\frac{\delta\tau}{x^2+\delta\tau^2}$ as a Lorentzian approximation for the delta function, with $\delta\tau=0.005$.
In Fig.~\ref{fig:density_of_states_function}(d) the value of $U$ as a function of the chemical potential is shown. 
We use these quantities to define a function analogous to Eq.~\eqref{eq:bcs_ord_par} as
\begin{equation}
    \delta_1^{}(\mu)=a e^{\frac{-b}{U(\mu)\Sigma(\mu)}},
    \label{eq:fit_1}
\end{equation}
and fit the coefficients $a$ and $b$ by minimizing the function $\chi^2=\sum_i \left(\Delta_1(\mu_i)-\delta_1(\mu_i)\right)^2$ as a function of the parameters. 
The result of the fit is shown in Fig.~\ref{fig:density_of_states_function}(a) with a purple line. 
As we can see the density of states is able to reproduce the peak of $\Delta_1$, but is unable to fit the oscillations and the the correct decrease as a function of the chemical potential. 
This is because Eq.~\eqref{eq:fit_1} does not contain information on the penetration depth of the surface states within the slab of the material, which is crucial to correctly reproduce the behaviour. 
In order to clarify this, we define an average value of the penetration function $\bar{f}_{z_0}$ as the weighted average
\begin{equation}
    \bar{f}_{y_{0}}(\mu)=\frac{\sum_n \int \frac{d^2 \mathbf{k}}{4\pi^2} f_{\mathbf{k}}^n(z_0) \mathcal{L}_{\delta\tau}(\tilde{\xi}_{\mathbf{k}}^n-\mu)}{\sum_n \int \frac{d^2 \mathbf{k}}{4\pi^2} \mathcal{L}_{\delta\tau}(\tilde{\xi}_{\mathbf{k}}^n-\mu)}.
    \label{eq:penetration_function_average}
\end{equation}
Replacing the density of states by the product of $\Sigma$ and Eq.~\eqref{eq:penetration_function_average} in Eq.~\eqref{eq:fit_1} we obtain the blue line in Fig.~\ref{fig:density_of_states_function}(a). 
Including the information on the penetration of the surface states results in a better agreement between the fit and the self-consistent data. 
We conclude that the density of states by itself cannot explain the superconductivity on the surface, but needs to be complemented with the information on the penetration of the Fermi arcs within the slab.

\section{Analytical tight-binding model}\label{app:analytical}
\begin{figure}
\centering
\includegraphics[width=0.49\textwidth, trim={0.8cm 0 0 1cm}]{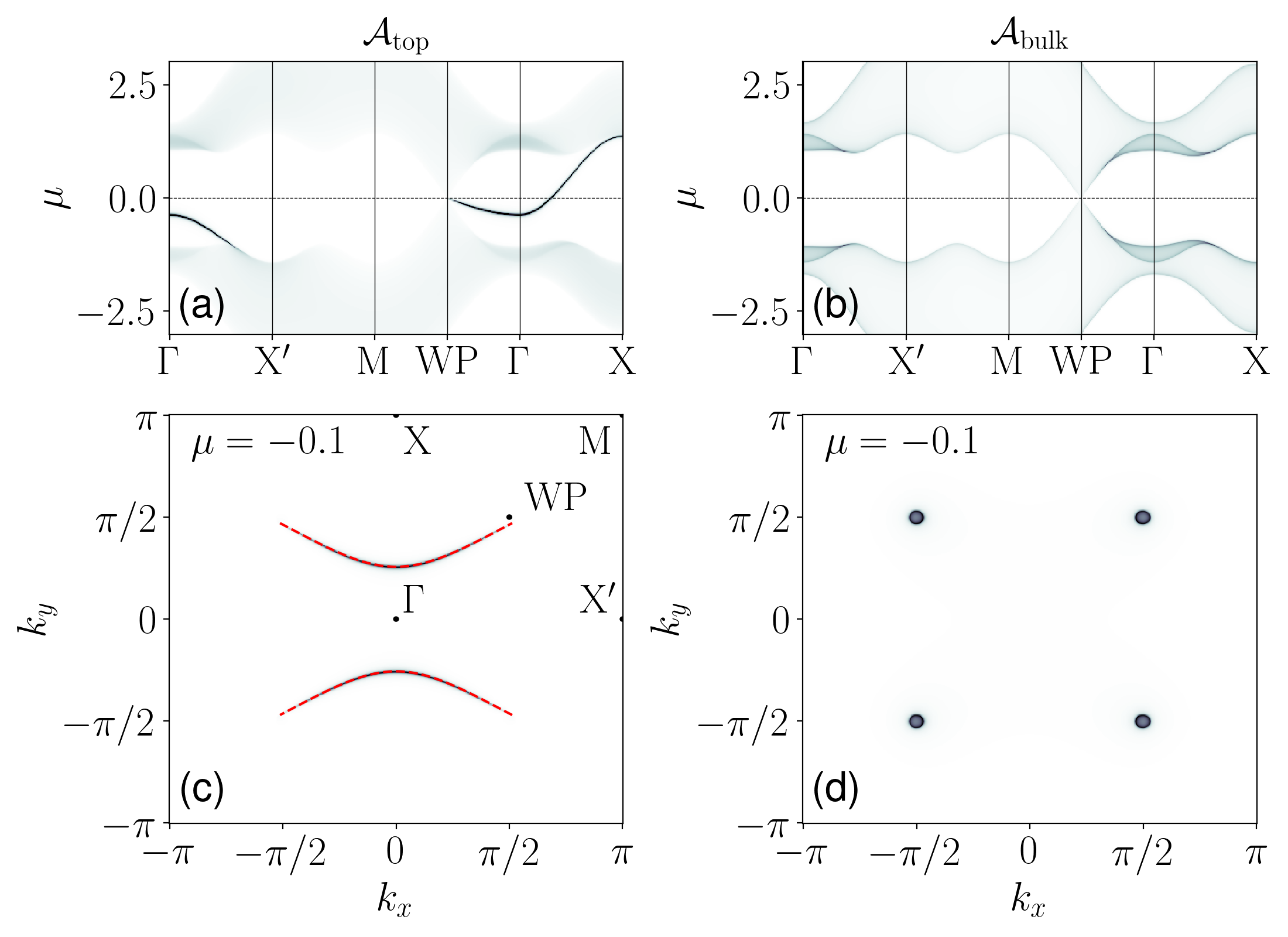}
\caption{
(a) Spectral function $\mathcal{A}_i$ of the top surface and (b) of the bulk of Hamiltonian Eq.~\eqref{eq:model_2}. 
The color maps of (a) and (b) are the same, and have been cut off at $\mathcal{A}^{\textrm{max}}=10$ to improve the contrast. 
(c) Spectral function of the surface and (d) of the bulk for a benchmark choice of the chemical potential $\mu=-0.1$. 
In dashed lines, the analytical eigenvalues matches the position of the Fermi arcs found with the numerical spectral function. 
Color maps of the two plots have been chosen to enhance the contrast. 
The parameters of the Hamiltonian are $t=1$, $\alpha=1.5$, $\gamma=\pi/6$, and the smearing of the Green's function $\delta=10^{-3}$.
}
\label{fig:spect_old_model}
\end{figure}

Here we discuss another tight-binding Hamiltonian of a time-reversal Weyl semimetal, one which can be analytically solved in the normal state, and thus can validate our numerical analysis. 
The Hamiltonian that we are taking into account is a generalization to a spinful Hamiltonian of the lattice model in Ref.~\cite{wawrzik2023berry}
\begin{equation}
    h(\mathbf{K})=
    \begin{pmatrix}
        h_{\uparrow\uparrow}(\mathbf{K}) & 0\\
        0 & h_{\downarrow\downarrow}(\mathbf{K})
    \end{pmatrix}
    \label{eq:model_2}
\end{equation}
where
\begin{widetext}
    \begin{align}
    \label{eq:hamUp} h_{\uparrow\uparrow}(\mathbf{K})=&-t\left(\cos(k_x)\tau_x +\cos(k_y)\tau_y - \sin(k_z)\tau_z \right)+ \alpha (1-\cos(k_z))[  \cos(\gamma)\tau_x +\sin(\gamma) \tau_y]-\mu\tau_0,
    \\
    h_{\downarrow\downarrow}(\mathbf{K})=& (h^{\uparrow\uparrow}(-\mathbf{K}))^*,
    \label{eq:hamDwn}
\end{align}
\end{widetext}
with matrices $\tau_i$ acting on the orbital degrees of freedom.
This model has the advantage that Weyl cones are always positioned at the four points $\mathbf{K}_{\textrm{Weyl}}=(\pm \pi/2,\pm(\kappa) \pi/2,0)$, with $\kappa=\pm 1$. 
In the following we fix the hopping amplitude to $t=1$, and note that $\alpha$ and $\gamma$ do not affect the location of the Weyl points, but only the spectrum of the Fermi arcs.
The eigenvalues of the Hamiltonian Eq.~\eqref{eq:hamUp} can be written analytically for both the bulk (since its diagonal blocks have the form $h=\mathbf{v}\cdot\boldsymbol{\tau}-\mu\tau_0$) and the $xy$ surface as
\begin{equation}
    \tilde{\xi}^e_{\pm}(\mathbf{k})=\pm(\cos(k_y) \cos(\gamma) - 
  \cos(k_x) \sin(\gamma)),
  \label{eq:sur_model2}
\end{equation}
within the domain defined by the expression
$|r(\mathbf{k})|\leq |\alpha|$,
with $r(\mathbf{k})=\alpha - \cos(k_x) \cos(\gamma) - 
  \cos(k_y) \sin(\gamma)$.
The spectral function for the surface and the bulk, together with the analytical Fermi surface are shown in Fig.~\ref{fig:spect_old_model}.
The sign $\pm$ in Eq.~\eqref{eq:sur_model2} refers to the bottom and top surfaces of a finite slab. For our purpose we will focus only on the top surface of a slab, which corresponds to the negative sign.

We write analytical expressions for the surface eigenstates $\ket{e_{\mathbf{k}}^n}$, after verifying that the ansatz 
\begin{equation}
    \ket{e^n_{\mathbf{k}}}=\sum_{ lz}\Tilde{A}_{\rho{_{\mathbf{k}}}}\; c_l \left[ \rho_{\mathbf{k}}^{z+1} - (\rho_{\mathbf{k}}^{z+1})^* \right] \ket{c_{\mathbf{k}lz}}
\end{equation} correctly satisfies the Schr\"odinger equation, where $\Tilde{A}_{\rho{_{\mathbf{k}}}}$ is the normalization of the vector, $c_l$ a coefficient that depends only on the orbital degrees of freedom and $|\rho_{\mathbf{k}}|<1$. 
The value of $\rho$ can be found by using the eigenvalues Eq.~\eqref{eq:sur_model2}, solving the Schr\"odinger equation with the ansatz above. 
We obtain
\begin{equation}               
 \rho_{\mathbf{k}}=\frac{r(\mathbf{k})\pm\sqrt{1-\alpha^2+r(\mathbf{k})^2}}{1+\alpha},
\end{equation}
where, again, $\pm$ refers to the bottom and top eigenvalues.
We can get the expression of $\Tilde{A}_{\rho{_{\mathbf{k}}}}$ by normalizing the eigenvectors, obtaining
$|\Tilde{A}_{\rho{_{\mathbf{k}}}}|^2=1/G_{\mathbf{k}}$, where
\begin{equation}
    G_{\mathbf{k}}=\left(\frac{2|\rho_{\mathbf{k}}|^2}{1-|\rho_{\mathbf{k}}|^2}-\frac{\rho_{\mathbf{k}}^2}{1-\rho_{\mathbf{k}}^2}-\frac{\rho_{\mathbf{k}}^{*2}}{1-\rho_{\mathbf{k}}^{*2}} \right).
\end{equation}
In this way, we obtain $f_{\mathbf{k}}(z_0)=T_{\mathbf{k}}/G_{\mathbf{k}}$, where
\begin{equation}
    T_{\mathbf{k}} = \Big(  \frac{2|\rho_{\mathbf{k}}|^2}{1-|\rho_{\mathbf{k}}|^2e^{-1/z_0}} - \frac{\rho_{\mathbf{k}}^2}{1-\rho_{\mathbf{k}}^2 e^{- 1/z_0}}
     - \frac{\rho_{\mathbf{k}}^{*2}}{1-\rho_{\mathbf{k}}^{*2}e^{- 1/z_0}} \Big).
\end{equation}
We can now exploit the fact that the BdG Hamiltonian can be decoupled in two distinct blocks since there is no spin-mixing in the single particle Hamiltonian in order to further simplify the problem. 
From the expression of the free energy in the main text we derive the self-consistent expression of the coupling strength $U$ of the bulk from the equation
\begin{equation}
    1=\frac{U}{2}\sum_{\pm}\int \frac{d^3 k}{(8 \pi^3)}  \frac{1}{\Lambda_\pm}\left( 1-\frac{2 e^{-\beta \Lambda_\pm}}{1+e^{-\beta \Lambda_\pm}}\right),
    \label{self-Delta0}
\end{equation}
where now $\pm$ refer to the two positive BdG eigenvalues $\Lambda_{\pm}=\sqrt{(\Xi_{\mathbf{k},k_z}\pm\mu)^2+U^2\Delta_0^2}$, where $\Xi_{\mathbf{k},k_z}$ refers to the eigenvalues of the bulk Hamiltonian without taking into account the chemical potential, and where we already summed over the two degenerate BdG eigenvalues due to the two spin-branches. 
\begin{figure}
\centering
\includegraphics[width=0.49\textwidth]{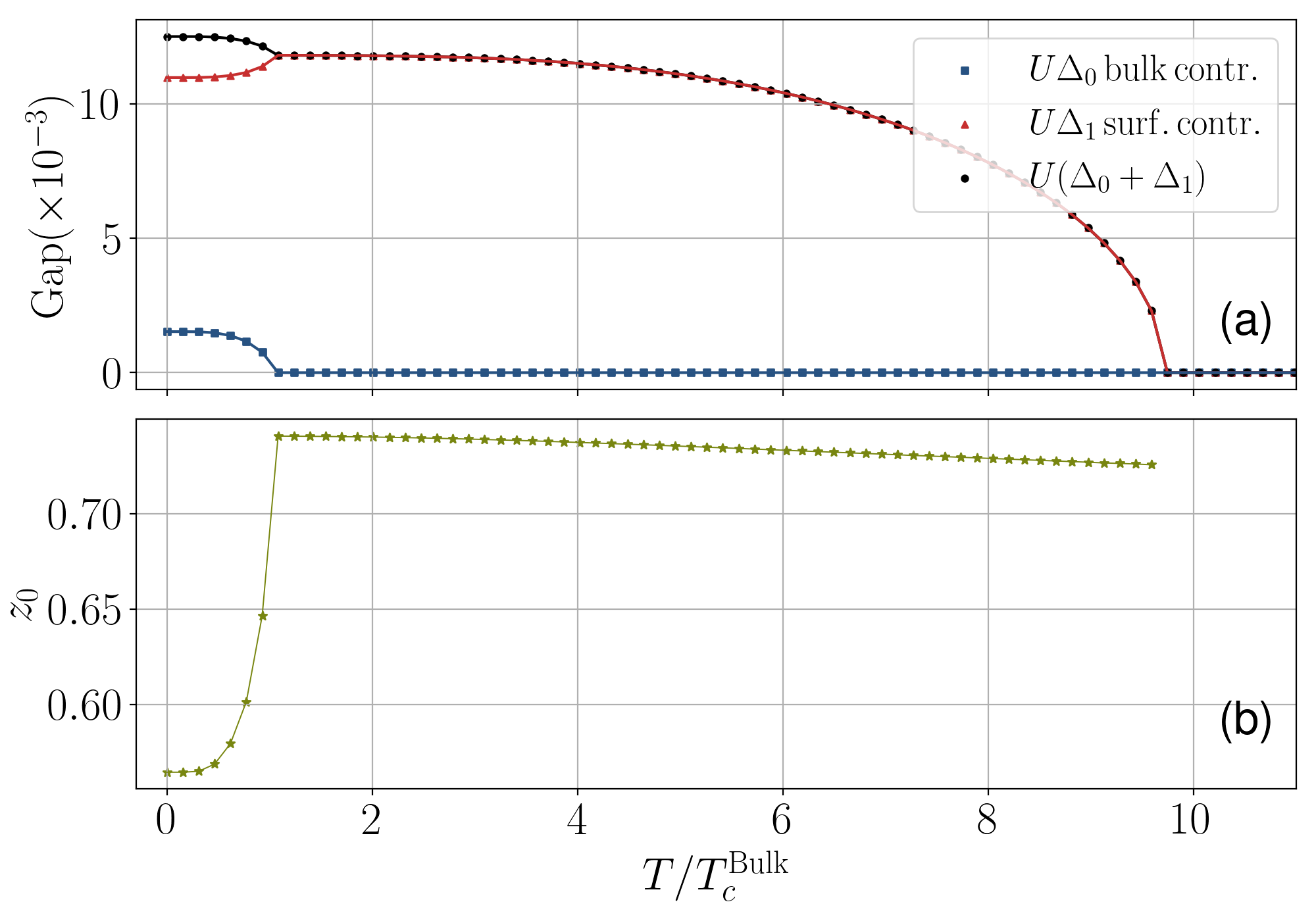}
\caption{
(a) Temperature dependence of the order parameter contributions. The blue, red, and black points correspond to the energy gap for the bulk ($g_0=U\Delta_0$), the surface ($g_1=U\Delta_1$) and for the total order parameter on the very first layer ($g_s=U(\Delta_0+\Delta_1)$), respectively.  
(b) Penetration depth of the surface order parameter as a function of the temperature. 
Above the critical temperature of the surface, $z_0$ loses meaning. 
All temperatures are rescaled to the bulk critical temperature $T_c^{\textrm{Bulk}}$.
}
\label{fig:model_old_ord_param}
\end{figure}

Using a value of U that opens a gap withing the bulk of $g_0=10^{-3}$, we can perform the self-consistent calculation of the surface free energy with the analytical value of the eigenvalues and the penetration function $f_{\mathbf{k}}(z_0)$. 
The results are collected in Fig.~\ref{fig:model_old_ord_param}. 
In this case we obtain an order parameter on the surface which is more than 10 times higher than the one in the bulk, as in the PtBi$_2$ case, showing that such an enhancement of the critical temperature is within the reach of an explanation based on the surface states as we propose here.

\section{Generalized variational ansatzes}\label{app:more_ansatzes}
The main advantage of the method we introduce in this work is that it is easily scalable to arbitrarily large systems; a direct numerical solution for the self-consistency of a spatially varying order parameter would otherwise be quite challenging to obtain, especially when multiple degrees of freedom and a large number of layers is considered. The principle by which we circumvent the difficulty is to incorporate the space dependence in a small number of variational parameters which are \textit{intensive}, independent of the size of the system. The main advantage that such a method offers is that it can be easily improved in precision by constructing generalized variational ansatzes for the order parameter. In this section, we perform such a test for the tight-binding model introduced in App.~\ref{app:analytical}.

Throughout the main text and our Supp. Mat., we have restricted to an exponential ansatz for the spatially varying part of the order parameter $\Delta=\Delta_0+\Delta_1 e^{-z/z_0}$. A simple non-exponential behavior that can be constructed involves an additional modulation by means of an oscillating factor $\Delta=\Delta_0+\Delta_1 e^{-z/z_0}\cos(\zeta z)$. Such oscillations might not be entirely unexpected, since the individual eigenstates generally exhibit a combined exponential and oscillating behavior in space~\cite{benito2019surface}. To test for this more general ansatz for the order parameter, we follow the same strategy as in the main text with an additional variational parameter $\zeta$.
\begin{figure}
    \centering
    \includegraphics[width=0.49\textwidth]{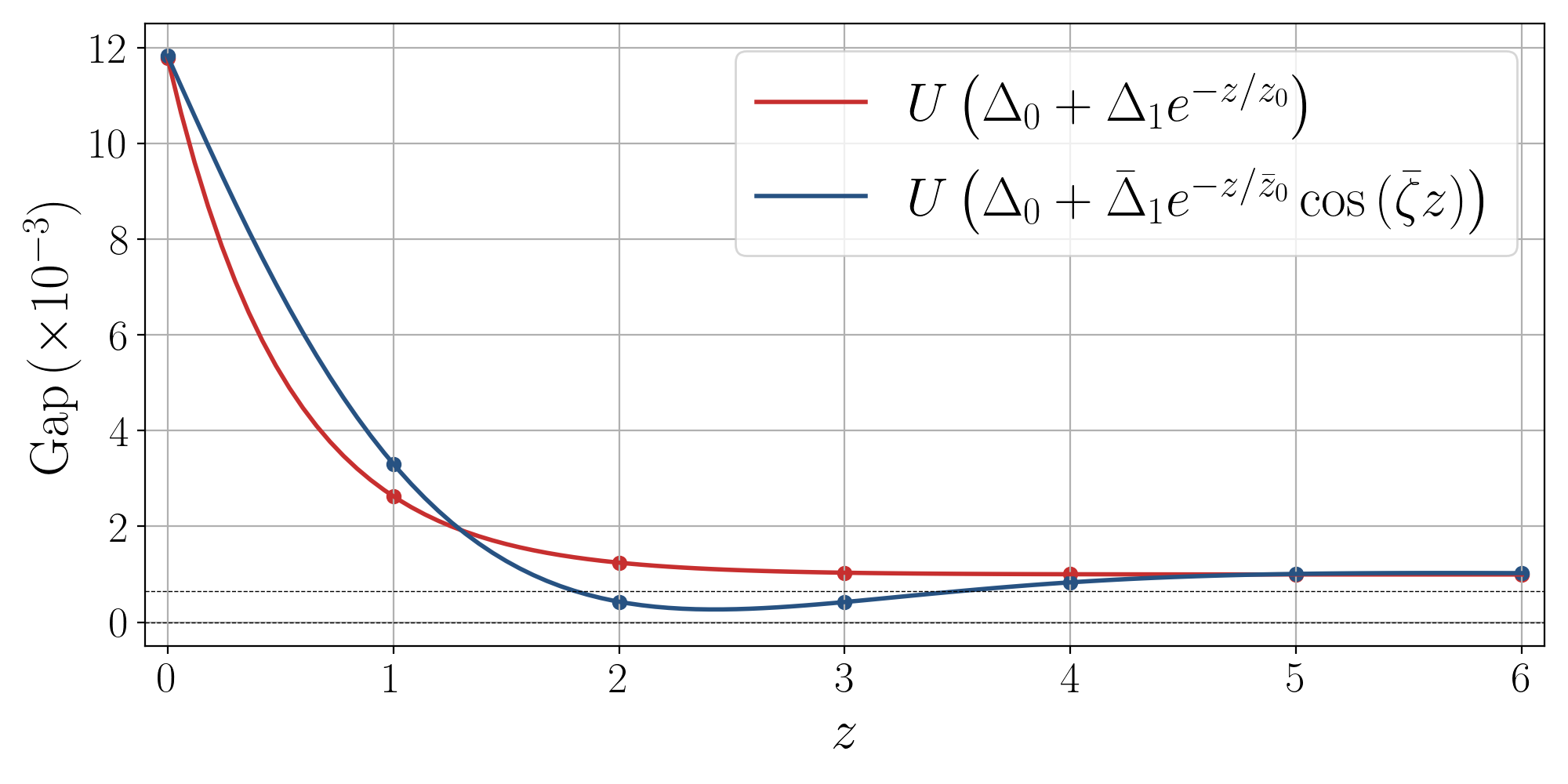}
    \caption{Superconducting gap as a function of the layer position using (red) the exponentially decaying and (blue) an exponentially and oscillatory decaying surface order parameters at zero temperature. The value that minimizes the free energy are $\Delta_1=7.17\times10^{-3}$, and $z_0=0.56$ for the first model (red) and $\bar{\Delta}_1=7.17\times10^{-3}$, $\bar{z}_0=0.98$ and $\bar{\zeta}=0.92$ for the second model (blue). The values of $U$ and $\Delta_0$ refer to the minimum found in App.~\ref{app:analytical}.}
    \label{fig:Oscillating_param}
\end{figure}

Fig.~\ref{fig:Oscillating_param} shows the minimum-free-energy solution for the order parameter at zero temperature. We compare the generalized solution, in which we minimize the surface free energy over $\Delta_1$, $z_0$, and $\zeta$, with the simple exponential ansatz, in which $\zeta$ is fixed to $0$ and the other parameters $\Delta_1$ and $z_0$ minimize the free energy. We find that a slow oscillation that modulates the spatial variation is indeed preferred, yet the qualitative picture remains the same as in the simple exponential ansatz. This retroactively justifies our results in the main text at the qualitative level; on the other hand, it also makes it clear that the order parameter in a realistic model can penetrate with a non-trivial structure within the surface layers of the crystal. The method we propose here offers a systematic way to obtain precise theoretical predictions for this structure, which is encoded in the variational parameters of the ansatz.

\bibliographystyle{unsrt}
\bibliography{Bib.bib}

\end{document}